\documentstyle[osa,aplop,manuscript]{revtex}

\begin{document}

\title{Discrimination between Nuclear Recoils and Electron Recoils by 
Simultaneous Detection of Phonons and Scintillation Light}

\author{ P.~Meunier,
M.~Bravin,
M.~Bruckmayer,
S.~Giordano,
M.~Loidl,
O.~Meier,
F.~Pr\"obst,
W.~Seidel,
M.~Sisti,
L.~Stodolsky,
S.~Uchaikin,
and L.~Zerle.}

\address{ Max Planck Institut f\"ur Physik,  F\"ohringer Ring 6, M\"unchen, 
Germany.
  }

\maketitle

\begin{abstract}
We have developed a detector, consisting of a
cryogenic calorimeter with a scintillating crystal as
absorber, and a second calorimeter for the detection of the 
scintillation light, both
operated at 12\,mK.
Using a CaWO$_4$ crystal with a mass of 6\,g as scintillating absorber, 
we have achieved a discrimination of nuclear recoils against
electron recoils with a suppression factor of 99.7\% at energies
above 15\,keV. This novel method will be applied for 
background rejection in the
CRESST (Cryogenic Rare Event Search with Superconducting Thermometers)  
experiment looking for dark matter Weakly Interacting Massive Particles
(WIMPs).

\begin{center}
{\copyright\ American Institute of Physics, 1999.}
\end{center}
\end{abstract}

Direct dark matter WIMPs (Weakly Interacting Massive Particles)
  searches are looking for nuclear 
recoil events caused by
WIMPs scattering off a nucleus in a detector. 
The kinetic energy of a recoiling nucleus due to a WIMP interaction is expected
to be of order of a few tens of keV.
Such recoils have a low ionization and scintillation efficiency  and
are therefore hard to detect in conventional detectors. 
In contrast, cryogenic calorimeters
are fully sensitive to nuclear recoils and can achieve much lower
energy thresholds~\cite{1,2,3} making them the most promising
detectors for a dark matter search. 

Since these dark matter events are expected
to be very rare, such experiments have to be carefully shielded against cosmic
radiation and ambient radioactivity. The  sensitivity of these experiments is
finally determined by the residual background, dominated by $\beta$, $\alpha$
and $\gamma$ emissions
from radioactive contaminations of the detectors and the surrounding 
materials.
  In contrast to WIMP interactions, this
background deposits energy in the detector mostly by electron recoils, 
which are characterized
by a higher ionization or scintillation efficiency than nuclear recoils. 
Thus the simultaneous measurement of ionization and phonons or scintillation
light and phonons can be used to discriminate 
between electron and nuclear recoils, and can therefore 
suppress this
background, leaving over nuclear recoil events caused by ambient neutrons
and WIMPs. 
The use of a cryogenic calorimeter in conjunction with conventional
ionization or scintillation measurement can provide high sensitivity together
with excellent discrimination capability. The simultaneous measurement of
ionization and phonons for a dark matter search has been developed and
successfully implemented in the CDMS (Cryogenic Dark Matter Search) 
experiment~\cite{3} and is also being used
in the EDELWEISS (Experience pour DEtecter Las WImps En Site Souterrain) 
experiment~\cite{4}. In view of achieving background
suppression in the CRESST (Cryogenic Rare Event Search with Superconducting 
Thermometers)  experiment - which relies on dielectric absorbers - 
\cite{1,2}, we have characterized several scintillators at millikelvin
temperatures. Since this kind of background rejection was
proposed~\cite{5,6}, it had only been possible to discriminate between
$\alpha$-particles and photons~\cite{7,8}. We report here simultaneous
measurement of scintillation light and phonons and demonstrate a clear
discrimination between electron and nuclear recoils down to 10 keV energy.

In order to obtain a good background suppression 
it is important to find dielectric crystals with good scintillation properties
at low temperatures.

For measuring different scintillators, we prepared
the scintillator holder in such a way that it was possible to remove and 
exchange the
scintillating crystal without any change in the light detector and in the
geometrical set-up.                                              
As light detector, we used a cryogenic calorimeter consisting of  a sapphire
substrate (10x20x0.5 mm$^3$) that was 
coated on one 10x20 mm$^2$ surface with silicon to improve the light absorption and had a
tungsten superconducting phase transition thermometer
deposited on the opposite surface. 
This set-up was mounted in a dilution refrigerator and cooled to about 12\,mK,
which is the operating temperature of the light detector.
An $^{241}$Am source was mounted inside the cryostat, irradiating the
scintillator with 5.5\,MeV $\alpha$ particles and 60\,keV photons.
For energy calibration, the light detector was irradiated with a 
$^{55}$Fe source emitting 5.9\,keV and 6.5\,keV X-rays. 
The scintillators  measured so far are BGO (Bi$_4$Ge$_3$O$_{12}$), BaF$_2$,
PbWO$_4$ and CaWO$_4$. We have found that they all work sufficiently well 
at mK temperatures.  Only the BGO crystal changed its properties after a 
few cooling cycles. 

In table\,I the energies of the detected light for the measured
scintillators are shown. These numbers are the detected energies and are not
corrected for light collection efficiencies which vary for different
scintillators because of their different refraction indices and light
wavelenghts. In BGO the detected light created by an $\alpha$ particle was a
factor 2.2 smaller than for a photon of the same energy. For CaWO$_4$ this
factor was 3.6.  

Since the scintillation effect in tungstates is connected to the
[WO$_4$$^{2-}$]-ion, 
presumably also other tungstates may work. Molybdates are other promising
candidates for scintillation detectors at mK temperatures. We will test these
materials in the near future.

Among the measured crystals, we chose CaWO$_4$ as a scintillator for 
the first simultaneous measurement of scintillation light and phonons. 
              
Fig.~1 shows the set-up. It consists of two independent detectors:
a 6g CaWO$_{4}$ scintillating crystal with 
a tungsten superconducting phase transition thermometer,
and  a second calorimeter placed next to it to 
detect the scintillation
light. The scintillator  has been surrounded by three aluminum
mirrors in order to improve the photon collection efficiency. With this set-up
an equivalent light energy of 480~eV has been measured 
for 60 keV $\gamma$-ray interactions in
CaWO$_4$.

Both detectors were operated at about 12mK.
 The CaWO$_4$ crystal
was irradiated with 122~keV and 136~keV photons from a $^{57}$Co source
and simultaneously with electrons from a $^{90}$Sr $\beta$-source.
The two sources contributed about equally to the count rate.
The photon lines were used for the energy calibration (in both
detectors). The trigger is provided by the phonon detector.

The left plot in fig.~2 shows a scatter plot of the pulse
height observed in the
phonon detector versus
the pulse height observed in the light detector. A clear
correlation between
the light and phonon signals is observed. The right hand plot shows an
additional irradiation with neutrons from an
americium-beryllium source.  A second
line can be seen  due to  neutron-induced nuclear recoils. It
is to be observed that electron
and nuclear recoils can be clearly distinguished down to a
threshold of
10~keV.

The scatter plots  in fig.~3 show the ratio of the pulse
height in the light detector
to the pulse height in
the phonon detector  versus the pulse height in the phonon detector.
The lower band in the right plot is caused by nuclear
recoils while the upper band with the ratio around 1 is caused
by electron
recoils. From the two ratios a quenching factor of 7.4 can be
inferred. The quenching factor for an $\alpha$ 
particle interaction is different, is
approximately half of it, but it still allows to discriminate $\alpha$ from
neutron interactions.

The leakage of some electron recoils  into the nuclear recoil line
gives the
electron recoil rejection according to the quality factor of ref.~\cite{9}.
A detailed evaluation together with the data without neutrons, shown in
the left plot of fig.~3,  yields a rejection factor of 98\% in
the energy
range between 10\,keV and 20\,keV, 99.7\% in the range between 15\,keV and
25\,keV and better than 99.9\% above 20\,keV.
The energy spectrum measured with the phonon detector due to the irradiation
with electrons and photons is shown in fig.~4a. Besides the lines at
122\,keV and 136\,keV there are two lines at 63\,keV and 55\,keV caused by the
escape of $K_\alpha$ and $K_\beta$ tungsten X-rays, excited by the 122\,keV
line in the CaWO$_4$ crystal. 
The corresponding energy spectrum measured in the light detector is shown
in figure 4b.

The residual background which can't be rejected with this method at the present
state of art consists of 
neutron events.
Other events that can have the same signature of a WIMP event are 
nuclear recoils induced by an $\alpha$ decay taking place inside the
detector very 
close to the surface so that the $\alpha$ particle can escape with an energy
release in the crystal lower than a few tens of keV. 
Also the interaction                 
of a nucleus emitted by the shield surface surrounding 
the detector as a consequence
af an $\alpha$ decay taking place at the surface of this material can have the
same signature of an WIMP event.
This kind of background due to $\alpha$ emitter surface contamination can be 
rejected completely by surrounding the  
CaWO$_4$ crystal with a passive shield made 
of the same scintillating material.
In this way, whenever the recoil nuclei induced by the $\alpha$ decay 
excites the CaWO$_4$ crystal, the light detector can also detect 
the scintillation light of the
$\alpha$ particle , and therefore the event can be rejected.

The simultaneous measurement of light and phonons 
has several advantages over the
simultaneous
measurement of phonons and charge. In the charge measurement electrical contacts
always produce an unfortunate dead layer on the  surface which causes
surface events, especially electrons from outside, to leak into the nuclear
recoil band. As our measurements with electrons clearly show, this problem
does
not exist in the light detection.
The large quenching factor of the CaWO$_4$ gives a very effective
separation
of nuclear recoils from electron recoils. 
As opposed to charge measurement, light collection does not suffer
from
problems such as space charge build-up, field inhomogeneities or
phonons produced by drifting the charges.
Due to these advantages many of the effects, known from
charge and phonon measurement to cause leakage of electron recoils into the
nuclear recoil band, are absent.  As a result the background suppression
efficiency of the light-phonon
detection is excellent and it works equally well for photon and electron
backgrounds, thus avoiding particle dependent systematic uncertainties in
the discrimination.

The opportunity to employ
different scintillators with different target nuclei  
gives an unique handle for
understanding  and reducing backgrounds in dark matter searches.
Even the neutron
background, always considered to be the ultimate limitation for
such experiments, could be investigated by varying the target nuclei.

In summary, a cryogenic particle detector which measures simultaneously 
the scintillation 
light and the thermal signal has been developed. 
Four different scintillating crystals have been tested
as cryogenic detector absorbers, and they all showed to work adequately at low
temperature. Using a 6~g scintillating CaWO$_{4}$ crystal,
it has been       
demonstrated that this technique allows a discrimination between nuclear and
electron recoils with a suppression factor of 98\% in the energy range
between 10keV and 20keV, 99.7\% in the range between 15~keV 
and 25~keV and better
than 99.9\% above 20keV. 
Applying our present techniques~\cite{2} and optimizing the design 
we are very confident
to produce detectors with a mass of a few hundred grams and similar
performances.                           

This work was supported by the Max Planck Institute for Physics, 
and by the EC-Network Program "Cryogenic Detectors", Contract 
no.ERBFMRXCT980167.

\begin{figure}
\caption{Schematic view of the arrangement used for the simultaneous 
 light and phonon detection.} 
\end{figure}
\begin{figure} 
\caption{Pulse height in the light detector versus pulse height in the
  phonon detector. The scatter plot on the left side has been measured
  with an electron and a photon source, while a neutron source was added 
  to measure the right plot.}
\end{figure}
\begin{figure}
\caption{Ratio of the pulse height in the light detector to the pulse height 
  in the phonon detector versus pulse height in the phonon detector   
  for irradiation with photons and electrons (left), 
  and with photons, electrons and neutrons (right). Electron recoils are  
  in the upper band while nuclear recoils are in the lower band.} 
\end{figure}
\begin{figure}
\caption{a): The energy spectrum measured with the phonon detector;   
  the detector has been irradiated with a $^{90}$Sr $\beta$ source and a 
  $^{57}$Co $\gamma$ source and neutron source. 
  b):  The measured energy spectrum with the light detector;     
  the energy scale applies for the heat measurement.} 
\end{figure}

\begin{table}
\renewcommand{\thetable}{\Roman{table}}
\caption{Energies of the detected light for the measured
scintillators at a working temperature of 12 mK.}
\begin{tabular}{lllccccc} \hline
Crystal & Detected Light       &  Detected Light  \\ 
        &    Energy (keV)      & Energy (eV)      \\ 
        & for 5.5 MeV $\alpha$ &  for 60 keV $\gamma$   \\   \hline    
CaWO$_4$  & 5.2             &  210     \\
BGO       & 8.4             &  200     \\ 
PbWO$_4$  & 1.9             &  -       \\ 
BaF$_2$   & 2.1             &  -       \\ 
       \hline                    
\end{tabular}
\end{table}

\begin{references}     
\bibitem{1} M. Buehler, L. Zerle, F. Proebst, A. Rulofs, U. Schanda, W. Seidel,
C. Abmaier, N. E. Booth, C. Bucci, P. Colling, S. Cooper, F. v.Feilitzsch, P.
Ferger, G. Forster, A. Gabutti, C. Hoess, J. Hoehne, J. Igalson, E. Kellner, M.
Koch, M. Loidl, O. Meier, A. Nucciotti, U. Nagel, M. J. J. v.d. Putte, G. L.
Salmon, M. Sisti, L. Stodolsky, A. Stolovich, "Status and low background
consideration for the CRESST dark matter search",
Nucl. Instr. Meth. A {\bf 370} 237 (1996).
\bibitem{2} M. Sisti, C. Bucci, M. Buehler, S. Cooper, F. v.Feilitzsch, 
J. Hoehne, V. Joergens, M. Loidl, O. Meier, U. Nagel, F. Proebst, 
A. Rulofs, M. L. Sarsa, J. Schangal, W. Seidel, L. Stodolsky, A. Stolovits, L.
Zerle, "Performance of the CRESST detectors and status of the experiment",
proceedings of the 7th International Workshop on Low
Temperature Detectors LTD-7, August 1997, Munich, Germany. 
\bibitem{3} T. Shutt, N. Wang, B. Ellman, Y. Giraud-Heraud, C. Stubbs, P. D.
Barnes, A. Cummings, A. DaSilva, J. Emes, E. E. Haller, A. E. Lange, J. Rich,
R. R. Ross, B. Sadoulet, G. Smith, W. Stockwell, S. White, B. A. Young, D.
Yvon, "Simultaneous high resolution measurement of phonons and ionization
created by particle interaction in a 60 g Germanium crystal at 25 mK", 
Phys Rev. Lett. {\bf 69}, 3531 (1992).
\bibitem{4} G. Chardin, A. Benoit, L. Berge, B. Chambon, M. Chapellier, P.
Charvin, M. De Jesus, P. Di Stefano, D. Drain, L. Dumoulin, C. Goldbach, A.
Juillard, D. Lhote, J. Mallet, S. Marineros, L. Miramonti, L. Mosca, X. F.
Navick, G. Nollez, P. Pari, C. Pastor, S. Pecourt, R. Tourbot, D. Yvon,
"Preliminary results of the EDELWEISS experiment",
Proc. of the 2nd Int. Workshop on the
Identification of Dark Matter, Buxton, England, 7-11 Sept.1998.
\bibitem{5} L. Gonzales-Mestres,
Proc. of the IV International
Workshop on Low Temperature Detectors for Neutrinos and
Dark Matter (LTD IV), Oxford, UK, Sept. 4-7 (1991). 
\bibitem{6}  J. S. Adams, S. R. Bandler, S. M. Brouer, R. E. Lanou, H. J.
Maris, T. More, G. M. Seidel, "Simultaneous calorimetric detection of rotons
and photons generated by particles in superfluid helium",
Phys. Lett. B {\bf 341} 189 (1995).      
\bibitem{7} A. Alessandrello, V. Bashkirov, C. Brofferio, C. Bucci, D. V. Camin,
O. Cremonesi, E. Fiorini, G. Gervasio, A. Giuliani, A. Nucciotti, M. Pavan, G.
Pessina, E. Previtali, L. Zanotti, "A scintillating bolometer for experiments
on double beta decay", Phys. Lett. B {\bf 420}, 109 (1998).
\bibitem{8} C. Bobin, I. Berkes, J. P. Hadjout, N. Coron, J. Leblanc, P. de
Marcillac, "Alpha/gamma discrimination with a $CaF_2(Eu)$ target bolometer
optically coupled to a composite infrared bolometer", 
Nucl. Instr. and Meth. A {\bf 386}, 453 (1997).
\bibitem{9} J. R. J. Gaitskell, P. D. Barnes, A. DaSilva, 
B. Sadoulet, T. Shutt, "The statistic of background rejection in direct
detection experiment for dark matter",
Nucl Phys. B. (Proc. Suppl.)    
{\bf 51} B, 279 (1996).
\end{references}
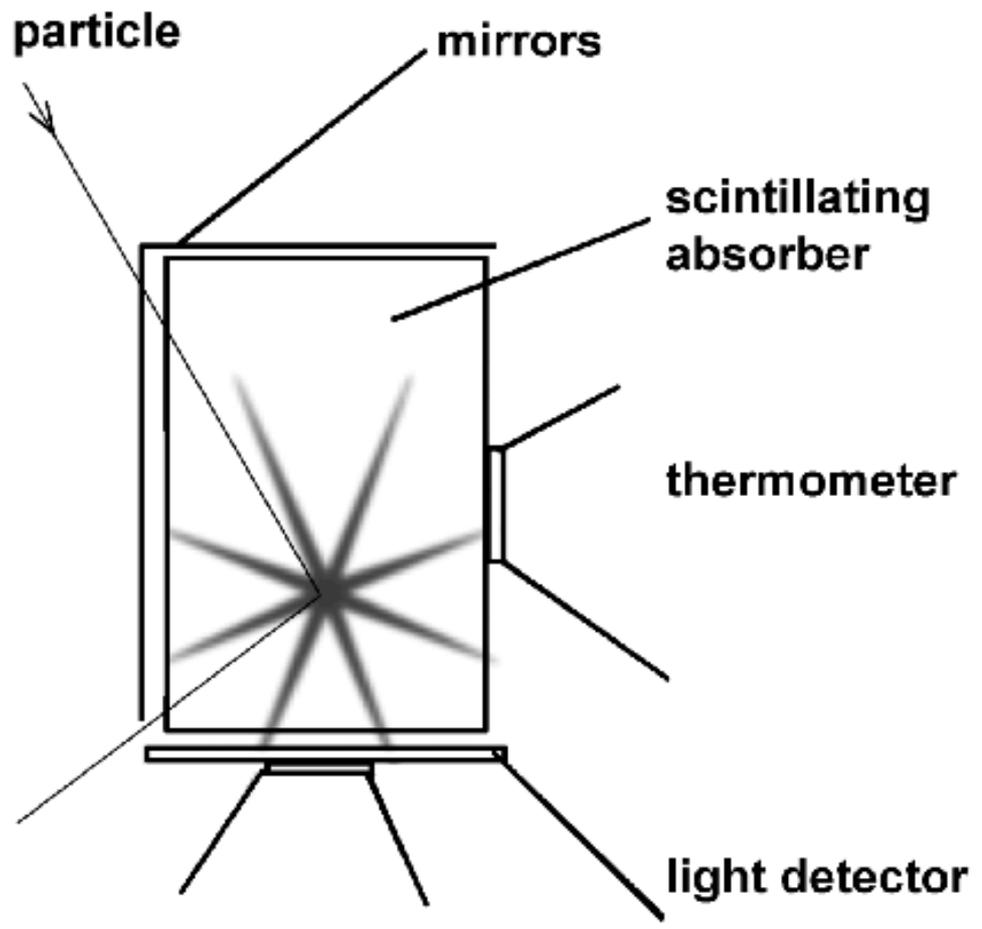
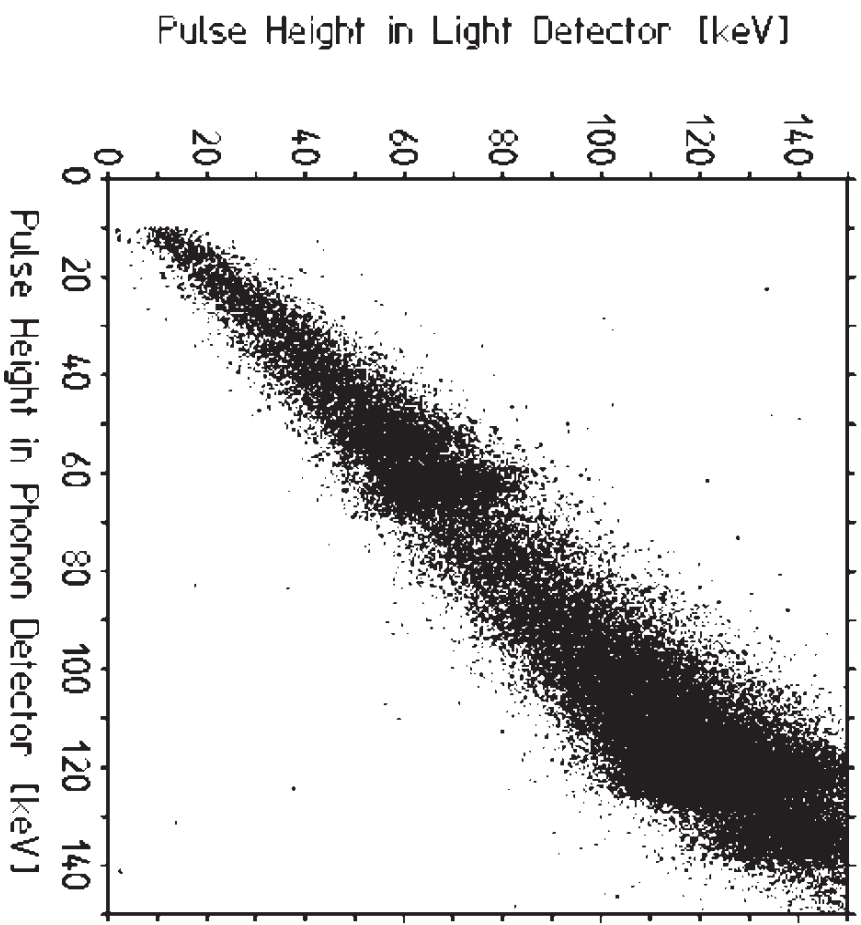
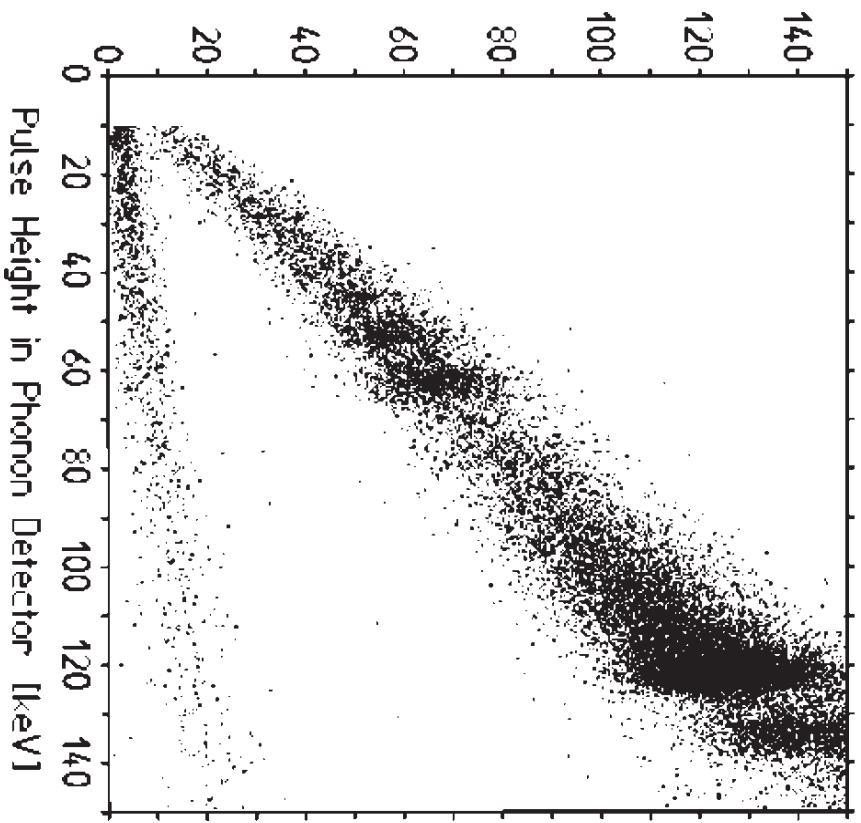
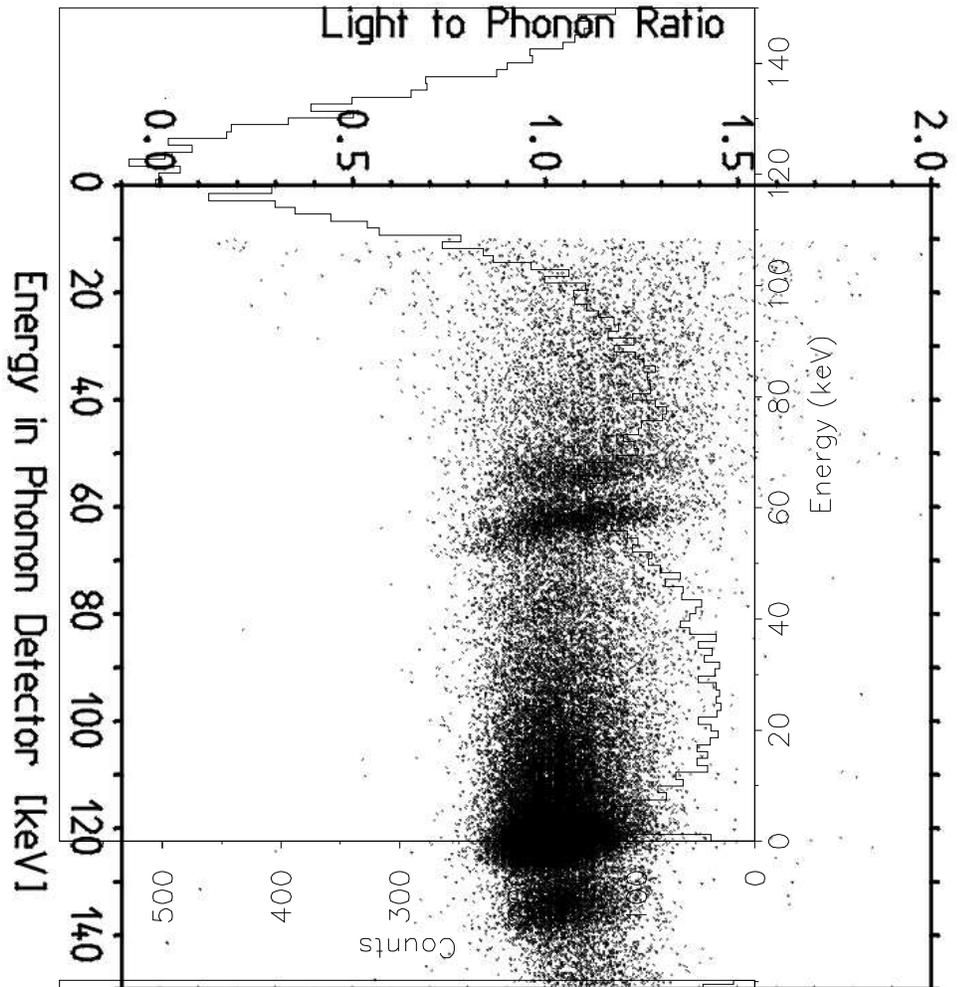
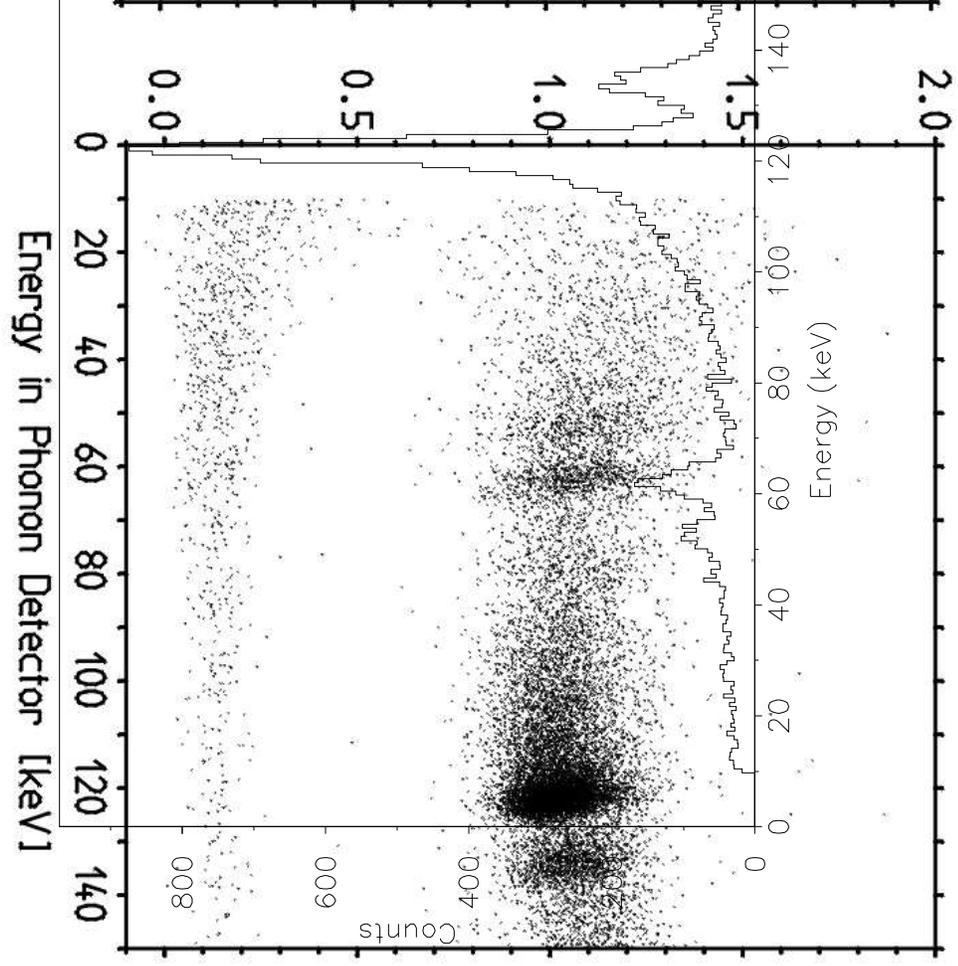
\end{document}